\documentclass[aps,pra,twocolumn,showpacs,longbibliography,superscriptaddress]{revtex4-1}
\usepackage[colorlinks=true, allcolors=blue]{hyperref}
\usepackage{amsmath}
\usepackage{amssymb}
\usepackage{braket}
\usepackage{color}
\usepackage{graphicx}
\usepackage{xcolor}

\newcommand{\la}{\langle}
\newcommand{\ra}{\rangle}
\newcommand{\da}{\dagger}
\newcommand{\vac}{|{\rm vac} \ra}

\newcommand{\cL}{\mathcal{L}}

\newcommand{\cD}{\mathcal{D}}

\newcommand{\cN}{\mathcal{N}}

\newcommand{\eps}{\varepsilon}
\newcommand{\Op}[1]{\hat{#1}}

\newcommand{\oH}{\Op{H}}

\newcommand{\oD}{\Op{D}}

\newcommand{\oS}{\Op{S}}

\newcommand{\oa}{\Op{a}}

\newcommand{\op}{\Op{p}}
\newcommand{\oq}{\Op{q}}

\newcommand{\tr}{\ensuremath{{\rm tr}}}
\newcommand{\diff}{\mathrm{d}}

\begin{document}
\title{Squeezed comb states}
\author{Namrata Shukla}
\email{namrata.shukla@mpl.mpg.de}
\affiliation{Max Planck Institute for the Science of Light, Erlangen 91058 , Germany}
\affiliation{%
	Institute for Quantum Science and Technology,
	University of Calgary, Alberta T2N 1N4, Canada}
\author{Stefan Nimmrichter}
\email{stefan.nimmrichter@uni-siegen.de} 
\affiliation{Naturwissenschaftlich-Technische Fakult{\"a}t, Universit{\"a}t Siegen, Siegen 57068, Germany}
\author{Barry C. Sanders}
\email{sandersb@ucalgary.ca}
\affiliation{%
	Institute for Quantum Science and Technology,
	University of Calgary, Alberta T2N 1N4, Canada}
\date{\today}

\begin{abstract}
Continuous-variable codes are an expedient solution for quantum information processing and quantum communication involving optical networks. Here we characterize the squeezed comb, a finite superposition of equidistant squeezed coherent states on a line, and its properties as a continuous-variable encoding choice for a logical qubit. The squeezed comb is a realistic approximation to the ideal code proposed by Gottesman, Kitaev, and Preskill 
\cite{GKP01}
which is fully protected against errors caused by the paradigmatic types of quantum noise in continuous-variable systems: damping and diffusion. This is no longer the case for the code space of finite squeezed combs, and noise robustness depends crucially on the encoding parameters. We analyze finite squeezed comb states in phase space, highlighting their complicated interference features and characterizing their dynamics when exposed to amplitude damping and Gaussian diffusion noise processes. We find that squeezed comb state are more suitable and less error-prone when exposed to damping, which speaks against standard error correction strategies that employ linear amplification to convert damping into easier-to-describe isotropic diffusion noise. 
\end{abstract}

\maketitle

\section{Introduction} 

Classical and quantum information is stored and accessed in discrete units, i.e.~bits and qubits~\cite{Sch95},
respectively,
but the actual physical encoding can be embedded in continuous, infinite-dimensional systems. Continuous-variable encoding of quantum information, in particular, may have a practical advantage in communication and computation implementations, given the readily available toolbox of linear optics and coherent states of light~\cite{WPG+12}. Moreover, the encoding of a finite set of distinct logical states in terms of a higher-dimensional system facilitates quantum error correction~\cite{Sho95}. 

Gottesman, Kitaev, and Preskill (GKP) introduced a continuous-variable code that represents quantum states of finite-dimensional Hilbert space by infinite `comb-like' superpositions of displaced position or momentum quadrature states in harmonic oscillator systems~\cite{GKP01}.
As quadrature eigenstates are inherently unphysical,
a realistic approximate GKP code based on squeezed coherent states was proposed for practical implementations.
The GKP proposal for realizing a cubic phase state, in particular, was analyzed in a case study~\cite{GS07}, showing that feasible levels of squeezing could not facilitate a close approximation to the ideal cubic phase state.

Despite practical limitations, superpositions of Gaussian wave packets with limited squeezing can serve as a viable encoding for quantum information processing in experiments~\cite{Travaglione02,Pirandola06,Vasconcelos10,Terhal16, Motes17}. 
Recently, physical realizations of GKP encoding with squeezed coherent states were achieved for a qubit in a trapped-ion experiment~\cite{Fluhmann2019} and with a superconducting microwave cavity~\cite{Campagne2020}. The use of GKP codes for universal fault-tolerant quantum computing on a protected code subspace was also investigated in broad survey studies~\cite{Albert18, Bourassa20,Terhal2020}.

In this paper, we study a realistic GKP encoding with finite resources based on \emph{squeezed comb states}: finite superpositions of \emph{teeth},
i.e., equidistant, distinct wave packets with a finite amount of squeezing. We characterize these states with the help of the Wigner-Weyl phase-space representation~\cite{Sch01,Leo13}, and we assess the impact of standard noise channels on the code space and on code errors. 
Whereas the detrimental influence of noise and the counter-acting error correction protocols are usually described in terms of discrete operations~\cite{Chuang97,Cochrane99,Llyod98,Braunstein98,Niset08,Ketterer16}, here we consider a more natural dynamical framework and focus on the stability of GKP-like encodings under continuous noise channels.
It turns out that a squeezed comb encoding grows more robust against amplitude damping noise with increasing teeth, whereas it becomes less robust against diffusion noise.

The outline is as follows.
In~\S\ref{sec:squeezed comb}, we introduce GKP encoding of a qubit in terms of squeezed comb states and we discuss their phase-space representation in terms of the Wigner function, which illustrates their intricate interference features.
In~\S\ref{sec:Noise channels}, we study the evolution of the squeezed comb state in the presence of two paradigmatic noise models: the amplitude damping channel describing pure loss of energy quanta to a zero-temperature bath, and the isotropic Gaussian noise channel describing pure diffusion resulting from a pure loss channel and the equivalent amount of linear amplification. Both cases can be treated analytically in phase space. We evaluate various figures of merit characterizing the sensitivity of the encoding to noise, including the state distinguishability that is a direct measure of code errors. Our findings suggest that GKP codes are more robust against the damping channel than against the diffusion channel. In~\S\ref{sec:conclusions}, we summarize our findings and conclude.

\section{Squeezed comb state}\label{sec:squeezed comb}

We first introduce the basis states for the finite GKP encoding of a qubit into superpositions of~$N$ equidistant squeezed coherent states along the position quadrature axis of a single-mode oscillator. Consider a Gaussian wave packet displaced by the coherent amplitude~$\alpha$ and squeezed to an amount characterized by the squeezing parameter $r$~\cite{LK87}. It can be obtained by applying first the squeezing operator $\oS(r)$ and then the displacement operator $\oD(\alpha)$ to the vacuum state $\vac$, with
\begin{align}
    \oD\left(\alpha\right)=\exp \left(\alpha^* \oa -\alpha \oa^\da \right), \, \oS\left(r\right)=\exp \left( r\frac{\oa^2 - \oa^{\da 2}}{2}\right).
\end{align}
We use the convention of dimensionless position and momentum quadratures defined via $\oa=(\oq +\text{i}\op)/\sqrt2$, such that the free Hamiltonian of the mode and the displacement operator become 
\begin{align} \label{eq:H}
    \oH &=\hbar\omega \left( \oa^\da \oa+\frac12 \right)=\hbar\omega \frac{\op^2+\oq^2}{2}, \\
    \oD (\alpha) &=\exp \left( \frac{\alpha^*-\alpha}{\sqrt2}\oq +\text{i}\frac{\alpha^*+\alpha}{\sqrt2}\op \right) \label{eq:D_pq},
\end{align}
which implies that position displacement of a wave function by $+q_0$ is represented by $\oD(-q_0/\sqrt2)$.
All of the following are formulated in the rotating frame with respect to $\oH$, in which the states do not evolve.

We now define the squeezed comb state encoding the computational basis of a qubit as a uniform superposition of equally spaced and equally squeezed coherent states on a line, the \emph{teeth} of the comb; the two basis states differ by a displacement of half the teeth spacing $d$,
\begin{eqnarray}\label{eq:squeezed comb}
    \ket{\bar0}&:=&\frac{1}{\sqrt{\cN}} \sum_{n=1}^N\oD\left( -\frac{\bar{q}_n}{\sqrt2} \right) \oS(r) |{\rm vac}\ra, \\
    \ket{\bar1}&:=&\oD\left(-\frac{d}{2\sqrt2} \right) \ket{\bar0} . \nonumber 
\end{eqnarray}
For minimal average energy, we choose the comb state representing the logical state $\ket0$ to be centered around the origin in phase space, $\sum_N\bar{q}_n=0$. The positions of the~$N$ teeth are then
\begin{equation}
    \bar{q}_n=\bar{q}_0 + nd, \, n=1,\dots,N, \, \bar{q}_0=-\frac{N+1}{2}d.
\end{equation}

As the teeth have an exponentially suppressed but finite overlap, the normalization factor $\cN$ in the above definition is
\begin{eqnarray}
    \cN &=&\sum_{n,m=1}^N\exp \left[-\text{e}^{2r} \frac{(\bar{q}_n-\bar{q}_m)^2}{4} \right] \nonumber \\
     &\approx& N + 2(N-1)\exp \left(-\text{e}^{2r} \frac{d^2}{4} \right), 
\end{eqnarray}
Here the last line shows the leading order correction in the limit $e^r d \gg 1$ of non-overlapping teeth, in which $\cN\mapsto N$. 
Another consequence of the overlap is that the two comb states are not perfectly orthogonal.
We find
\begin{eqnarray} \label{eq:orthog_pure}
    \la \bar0\ket{\bar1}&=&\frac{1}{\cN} \sum_{n,m=1}^N\exp \left[-\frac{\text{e}^{2r}}{4} \left(\bar{q}_n-\bar{q}_m - \frac{d}{2}\right)^2 \right] \\
    &\approx&\frac{2N-1}{N} \exp \left(-\text{e}^{2r} \frac{d^2}{16} \right), \nonumber
\end{eqnarray}
once again with the lowest-order term for almost non-overlapping teeth.

For large combs, the scalar product is mainly determined by $(r,d)$ and no longer depends much on~$N$. It is directly related to the distinguishability of the basis states and to code errors, as we discuss in~\S\ref{subsec:distinguishability}. 
The ideal GKP code (with perfectly orthogonal basis states) would be reached asymptotically in the limit of infinitely large squeezing and tooth number, $r,N\to \infty$.

We proceed to analyze the features of the encoding, employing the Wigner-Weyl phase space representation. The Wigner function for a given single-mode state $\rho$ is
\begin{equation}
\label{eq:Wignerqp}
	w(q, p)	=\frac{1}{2\pi}\int \text{d}x~\text{e}^{\text{i}px}	\bigg\langle q-\frac{x}{2}|\rho|q+\frac{x}{2}\bigg\rangle .
\end{equation}
Its marginals yield the state's position and momentum distribution,
\begin{equation}
    f(q) = \int \diff p \, w(q,p), \qquad  f(p) = \int \diff q \, w(q,p),
\end{equation}
respectively.

The comb-state Wigner function for $\rho=\ket{\bar0} \langle \bar0|$ is given analytically as
\begin{eqnarray}
\label{eq:wigner_sqcomb}
    w_{\bar0}(q,p) &=&\frac{\exp (-\text{e}^{-2r} p^2)}{\cN\pi} \sum_{n,m=1}^N\cos p (\bar{q}_n-\bar{q}_m)  \nonumber \\
    &&\times \exp \left[- \text{e}^{2r} \left( q - \frac{\bar{q}_n+\bar{q}_m}{2} \right)^2 \right]
\end{eqnarray}
This expression is real-valued, and the double summation separates the purely positive terms resulting from a classical mixture of the teeth ($n=m$) from the interference terms oscillating along the momentum axis ($N\neq m$). The other basis state is encoded with $w_{\bar1}(q,p)=w_{\bar0}(q-d/2,p)$. For the position marginals, we obtain
\begin{eqnarray}\label{eq:qmarginal_sqcomb}
    f_{\bar0}(q) &=&\frac{e^r}{\cN\sqrt{\pi}} \sum_{n,m=1}^N\exp \left[ - \text{e}^{2r} \left( q - \frac{\bar{q}_n+\bar{q}_m}{2} \right)^2 \right] \nonumber \\
    &&\times \exp \left[ - \text{e}^{2r} \left( \frac{\bar{q}_n-\bar{q}_m}{2} \right)^2\right] \\
    &\approx&\frac{e^r}{\cN\sqrt{\pi}} \left\{ \sum_{n=1}^N\exp \left[ -\text{e}^{2r} \left( q - \bar{q}_n\right)^2 \right] \right. \nonumber \\
    &&+\left.  2 \sum_{n=1}^{N-1} \exp \left[ -\text{e}^{2r} \left( q - \bar{q}_n - \frac{d}{2} \right)^2 -\text{e}^{2r} \frac{d^2}{4} \right]  \right\}, \nonumber 
\end{eqnarray}
and $f_{\bar1}(q)=f_{\bar0}(q-d/2)$, respectively. 

The approximation in the last two lines of Eq.~(\ref{eq:qmarginal_sqcomb})
gives the relevant contributions in the limit of almost non-overlapping teeth:
a sum of~$N$ individual Gaussian teeth and a sum over small Gaussian side peaks in between the teeth. The momentum marginal for both basis states is
\begin{equation}
\label{eq:pmarginal_sqcomb}
    f_{{\bar0},{\bar1}}(p)
    =\frac{\text{e}^{-r}}{\cN\sqrt{\pi}} \exp \left( - \text{e}^{-2r}p^2\right)
     U_{N-1}^2\left(\cos\frac{pd}{2}\right) 
\end{equation}
for
\begin{equation}
U_n(\cos\theta)
    =\frac{\sin[(n+1)\theta]}{\sin\theta}
\end{equation}
the Chebyshev polynomial of the second kind.
The marginal~(\ref{eq:pmarginal_sqcomb}) describes the Fraunhofer diffraction pattern emerging from a grating of~$N$ Gaussian slits, which reflects the nonclassical features of GKP encoding.

\begin{figure}
\centering
\includegraphics[width=0.95\columnwidth]{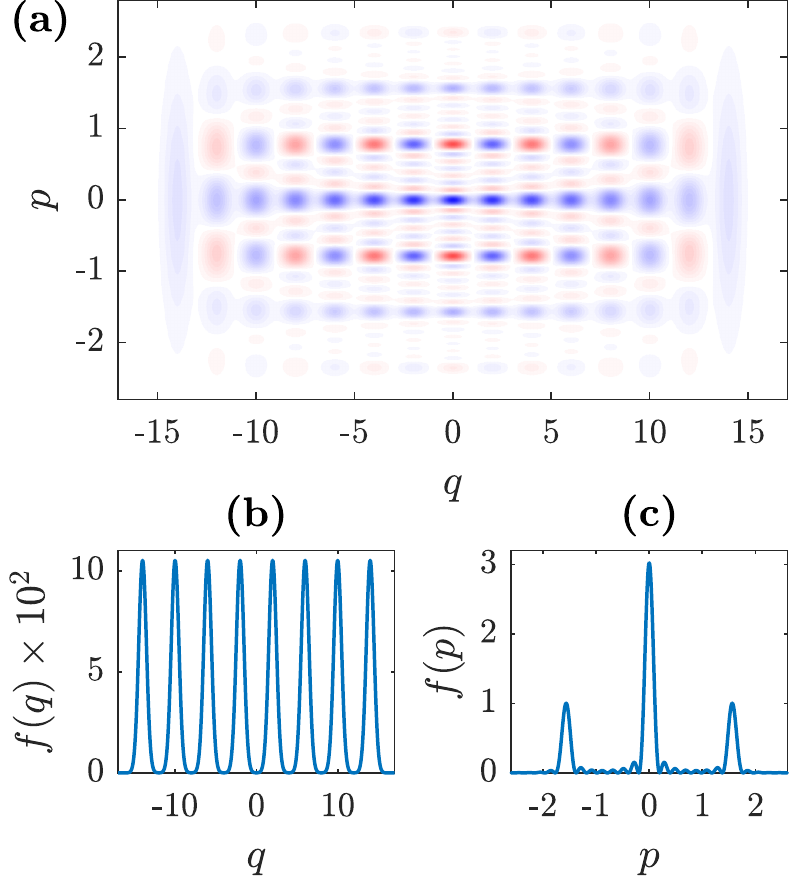}
\caption{(a) Wigner function of the squeezed comb state $\ket{\bar0}$, with $N=8$ teeth at spacing $d=4$ and squeezing $r=0.4$. The blue and red shades mark regions of positive and negative values, respectively. (b) Corresponding position marginal distribution. (c) Momentum marginal distribution.}
 \label{fig:Wignermarginal1}
\end{figure}

\begin{figure}
\centering
\includegraphics[width=0.95\columnwidth]{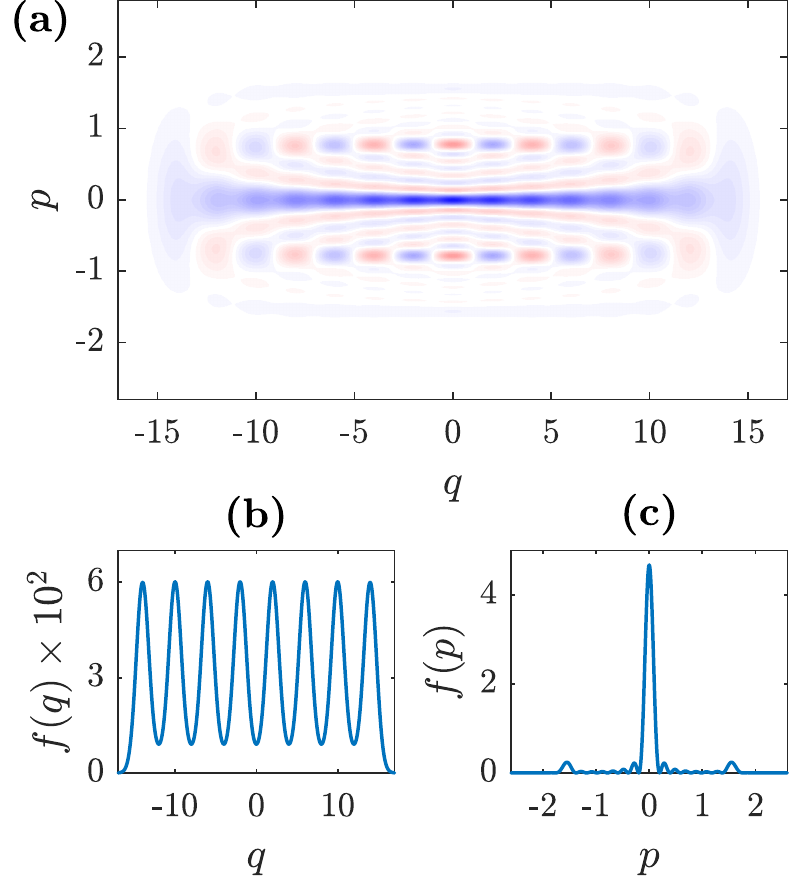}
\caption{%
(a) Wigner function of $\ket{\bar0}$, with $N=8$ teeth at $d=4$ and anti-squeezing $r=-0.1$. Blue and red indicate positive and negative values, respectively. (b) Position marginal distribution. (c) Momentum marginal distribution.}
 \label{fig:Wignermarginal2}
\end{figure}

Figures \ref{fig:Wignermarginal1} and~\ref{fig:Wignermarginal2} depict exemplary plots of the Wigner function and its marginals for the comb state $\ket{\bar0}$ with $N=8$ teeth at two different levels of squeezing.
Oscillatory fringe patterns with negativities appear along the vertical $p$ axis, centered around the $2N-1$ positions $q=(\bar q_n + \bar q_m)/2$ at $p=0$, as described by Eq.~\eqref{eq:wigner_sqcomb}.
The central fringe pattern at $q=0$ has the highest amplitude, as it comprises all teeth interfering in phase, and the amplitudes decrease symmetrically to both sides.
Decreased separation or squeezing along the position axis results in a greater overlap between the Gaussian teeth and a washed out interference pattern as in Fig.~\ref{fig:Wignermarginal2}.
Similar phase-space interference features are analyzed for quantum tetrachotomous states~\cite{SS19}. 

\section{Noise Channels}
\label{sec:Noise channels}
For the most common scenario in which GKP codes are realized with optical modes, and also for vibrational modes in a cold trapped-ion setting, the encoded states are most likely to be subject to amplitude damping~\cite{Chuang97,Cochrane99}, i.e.,~energy loss to an effectively zero-temperature bath. With the help of linear parametric amplification, the damping channel can be converted to pure diffusion noise, i.e.,~random isotropic and Gaussian-distributed displacement errors acting on the code in a finite time interval. Ideal GKP states were shown to be robust against this type of noise provided the time intervals, or average displacements, between discrete error correction steps are small. The errors could then be detected non-destructively and corrected by controlled displacements. 

In a realistic GKP encoding with a finite number of finite-sized teeth, however, the displacement errors caused by damping and diffusion processes cannot be perfectly suppressed and the code space is not stable. Reliable operation of GKP codes is therefore a matter of competing time scales: processing time versus characteristic error accumulation. In the following, we assess the dynamics of GKP qubit encoding under damping and diffusion noise, which can be formulated analytically in phase space.

The time evolution of a GKP state under the influence of amplitude damping is described by the usual dissipator,
\begin{equation}
\dot{\rho}=\cL \rho=\gamma \oa\rho \oa^\da-\frac{\gamma}{2}\left(\oa^\da \oa\rho+\rho \oa^\da \oa \right),\label{eq:dissipation}
\end{equation}
in the interaction picture with respect to the free Hamiltonian~\eqref{eq:H}. The corresponding Fokker-Planck equation for the Wigner function is
\begin{equation}
    \dot{w}(q,p;t)=\left[\frac{\gamma}{4}(\partial_p^2+\partial_q^2)+\frac{\gamma}{2}(\partial_q q+\partial_p p)\right]w(q,p;t),
\label{eq:FPE}
\end{equation}
which can be solved with the help of a Fourier transform between the Wigner function and its associated characteristic function. 

The solution to Eq.~(\ref{eq:FPE}) is a Gaussian convolution with rescaled arguments,
\begin{align}
    w(q,p;t) =&\int \frac{\diff q_0 \diff p_0\, \text{e}^{\gamma t}}{\pi (1-\text{e}^{-\gamma t})} w \left(q_0 \text{e}^{\gamma t/2},p_0 \text{e}^{\gamma t/2} \right)  \nonumber \\
    &\times \exp \left[ \frac{(q-q_0)^2 + (p-p_0)^2}{1-\text{e}^{-\gamma t}} \right].
\end{align}
For the two basis states, it yields Gaussians in the position and momentum quadratures with time-evolved width parameters,
\begin{eqnarray}
\sigma_q^2(t) &=& 1-\text{e}^{-\gamma t}+\text{e}^{-2r-\gamma t}, \\
\sigma_p^2 (t) &=& 1-\text{e}^{-\gamma t}+\text{e}^{2r-\gamma t},
\end{eqnarray}
and time-rescaled displacements,
\begin{eqnarray}
    w_{\bar0}(q,p;t) &=&\frac{\text{e}^{-p^2/\sigma_p^2(t)}}{\cN\pi \sigma_q(t) \sigma_p(t)} \sum_{n,m=1}^N\cos \left[ \frac{(\bar{q}_n-\bar{q}_m)p}{\sigma_p^2(t) \text{e}^{-2r+\gamma t/2}} \right] \nonumber \\
    &&\times \exp \left[ - \frac{1}{\sigma_q^2 (t)} \left( q - \text{e}^{-\gamma t/2} \frac{\bar{q}_n+\bar{q}_m}{2} \right)^2 \right] \nonumber \\
    &&\times \exp \left[ -\frac{1-\text{e}^{-\gamma t}}{\sigma_p^2(t)\text{e}^{-2r}} \left( \frac{\bar{q}_n - \bar{q}_m}{2} \right)^2\right],
\end{eqnarray}
and
\begin{equation}
    w_{\bar1}(q,p;t)=w_{\bar0} \left(q-\text{e}^{-\gamma t/2}\frac{d}{2},p;t \right).
\end{equation}

Systematic decay towards the vacuum caused by amplitude damping~\eqref{eq:dissipation} can be eliminated by adding a linear amplifier at the same rate~\cite{Albert18},
which converts the damping channel to a Gaussian diffusion channel (i.e., an effectively infinite-temperature bath).
The associated master equation is
\begin{eqnarray}
    \dot{\tilde{\rho}}=\tilde{\cL} \tilde{\rho}&=&\gamma \left[ \oa \tilde{\rho} \oa^\da+\oa^\da \tilde{\rho} \oa - \frac12\{ \oa^\da \oa+\oa \oa^\da , \tilde{\rho} \} \right] \nonumber \\
    &=&\gamma \left[ \oq \tilde{\rho} \oq+\op \tilde{\rho} \op - \frac12 \{ \oq^2+\op^2 , \tilde{\rho} \} \right]. \label{eq:gaussianNoise}
\end{eqnarray}

Mitigating the damping comes at the price of doubling the diffusion rate, but the noise that one needs to error-correct simplifies to isotropic random phase-space displacements, as described by the Fokker-Planck equation 
\begin{equation}
\partial_t \tilde{w}(q,p;t)=\frac{\gamma}{2} (\partial_q^2+\partial_p^2) \tilde{w}(q,p;t) .
\end{equation}
The latter is solved by a Gaussian convolution with linearly growing spread,
\begin{equation}
    \tilde{w}(q,p;t)=\int \frac{\diff q_0 \diff p_0}{2\pi \gamma t} w(q-q_0,p-p_0) \text{e}^{-(q_0^2+p_0^2)/2\gamma t}.
\end{equation}
For the two basis states, we arrive at
\begin{eqnarray}
    \tilde{w}_{\bar0}(q,p;t) &=&\sum_{n,m=1}^N\frac{\cos \left[ p(\bar{q}_n-\bar{q}_m)/(1+2\gamma t \text{e}^{-2r}) \right]}{\cN\pi \sqrt{(\text{e}^{2r}+2\gamma t)(\text{e}^{-2r}+2\gamma t)}}  \nonumber \\
    &&\times \exp \left[ - \frac{1}{\text{e}^{-2r}+2\gamma t} \left( q - \frac{\bar{q}_n+\bar{q}_m}{2} \right)^2 \right] \nonumber \\
    &&\times \exp \left[ -\frac{2 p^2+\gamma t  \text{e}^{2r}(\bar{q}_n - \bar{q}_m)^2}{2 (\text{e}^{2r}+2\gamma t) }\right],
\end{eqnarray}
and
\begin{equation}
\tilde{w}_{\bar1}(q,p;t)=\tilde{w}_{\bar0} \left(q-\frac{d}{2},p;t \right)
\end{equation}

\begin{figure}
\centering
\includegraphics[width=0.95\columnwidth]{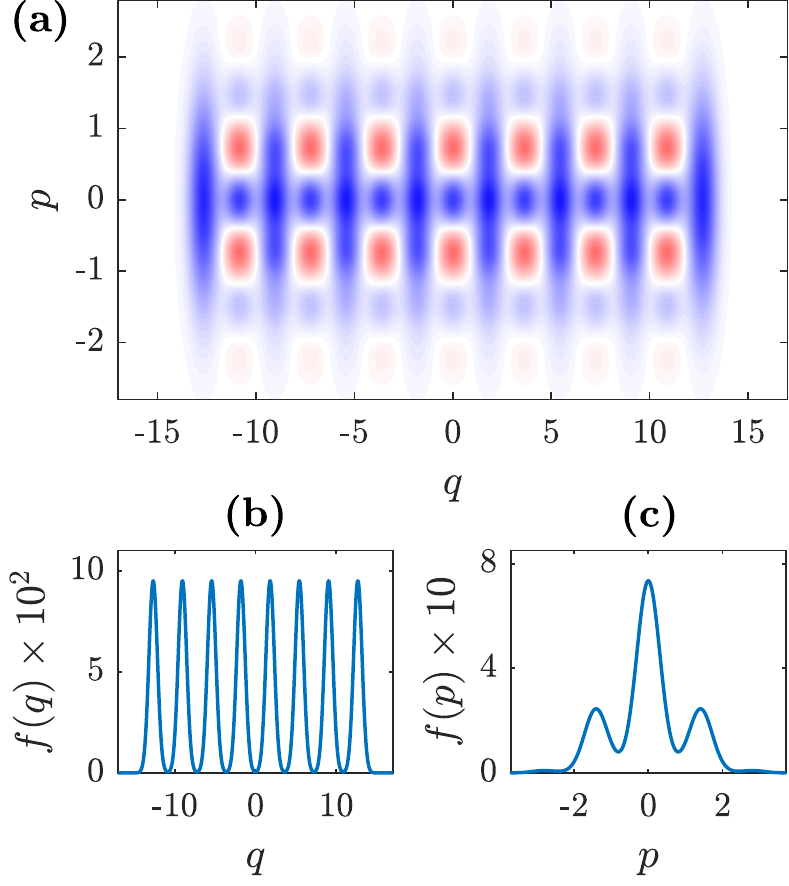}
\caption{Time-evolved Wigner function and marginals subject to amplitude damping at $\gamma t=0.2$, starting from the initial comb state of Fig.~\ref{fig:Wignermarginal1}. (a) Wigner function. Blue and red indicate positive and negative values, respectively. (b) Position marginal distribution. (c) Momentum marginal distribution.}
\label{fig:Wignermarginal1_t}
\end{figure}

\begin{figure}
\centering
\includegraphics[width=0.95\columnwidth]{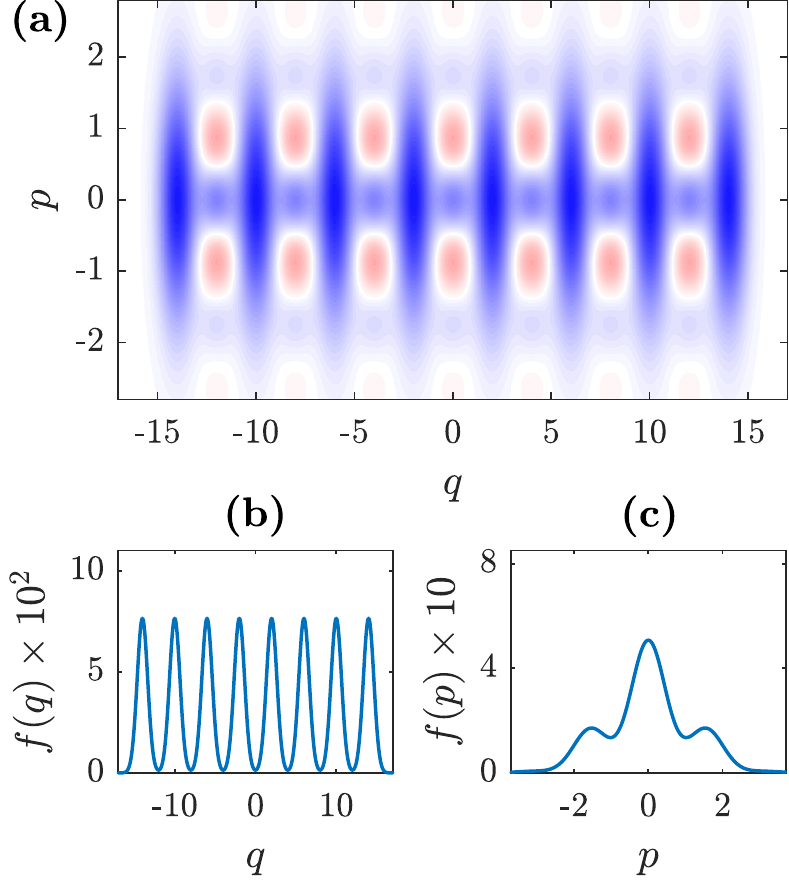}
\caption{Time-evolved Wigner function and marginals subject to Gaussian diffusion at $\gamma t=0.2$, starting from the initial comb state of Fig.~\ref{fig:Wignermarginal1}. (a) Wigner function. Blue and red indicate positive and negative values, respectively. (b) Position marginal distribution. (c) Momentum marginal distribution.}
 \label{fig:Wignermarginal2_t}
\end{figure}

Figures \ref{fig:Wignermarginal1_t} and \ref{fig:Wignermarginal2_t} show the time-evolved Wigner functions and marginals associated to the initial squeezed comb state $\ket{\bar0}$ with $(N,d,r)=(8,4,0.4)$ subject to amplitude damping and diffusion, respectively. Both cases are evaluated at $\gamma t=0.2$, which amounts to twice as much noise in the diffusion case. Most of the interference features are already washed out compared to the initial Wigner function plotted in Fig.~\ref{fig:Wignermarginal1}, but the leading-order diffraction peaks in the momentum marginal distribution (as well as the associated negative parts of the Wigner function) are still visible. At the evaluated time, the error probability for distinguishing the two basis states $\ket{\bar0}$ and $\ket{\bar1}$, initially at $1.0\%$, has grown to $2.7\%$ and $9.1\%$ for the damping case and for the diffusion case, respectively.

When code states are exposed to a damping or diffusion channel, they are affected in two ways: First, the states leave the code space spanned by the squeezed combs~\eqref{eq:squeezed comb}, and second, the basis states become less distinguishable leading to increased code errors. Both effects crucially depend on the comb parameters: the tooth number~$N$, the spacing $d$, and the squeezing parameter~$r$. In the following sections, we assess this parameter dependence in terms of time-evolved fidelities, orthogonality, and distinguishability between the code states. For practical purposes, the distinguishability discussed in \S\ref{subsec:distinguishability} is the most relevant quantity. It gives the lowest attainable error in measurements that discriminate the two computational states under noise.

\subsection{Fidelity}
\label{subsec:fidelity}

The fidelity between an initial code state and its time-evolved counterpart in the presence of a noise channel is a simple figure of merit that captures the departure from code space. Analytic results can be obtained, noticing that the expression for fidelity~\cite{Case08} between two states $\rho,\sigma$ reduces to the simple Hilbert-Schmidt scalar product if one of the states is pure,
\begin{equation}
F(\rho,\sigma)=\left(\tr \sqrt{\sqrt{\rho}\sigma\sqrt{\rho}}\right)^2 \xrightarrow{\rho=|\psi\ra\la \psi|} \la \psi | \sigma |\psi\ra=\tr (\rho\sigma).
\end{equation}
Here $|\psi\ra$ represents a comb state and $\sigma=\text{e}^{\cL t}(|\psi\ra\la \psi|)$ represents the same state after time $t$ under damping~\eqref{eq:dissipation}.
We denote the corresponding fidelity as $F_\psi (t)$, which can be conveniently expressed as an overlap integral of the respective Wigner functions,
\begin{equation}\label{eq:fidelitytowinger}
F_\psi (t)=2\pi \int \diff q \diff p\, w_\psi(q,p) w_\psi(q,p;t).
\end{equation}
From the initial $F_\psi(0)=1$ onwards, the fidelity will decay, and for the case of the damping channel, it will eventually reach the much lower final value $F_\psi(\infty)=|\la \psi \vac |^2$ in the limit $\gamma t \gg1$.

In order to stabilize the initial code space spanned by the pure comb states $|\bar{0}\ra,|\bar{1}\ra$ in a practical implementation, one would have to monitor and counteract already small changes of fidelity as quickly as possible.
A figure of merit for the required frequency of monitoring and stabilization operations would be the initial decay rate of the fidelity.
For the damping channel generated by~\eqref{eq:dissipation}, we arrive at
\begin{equation}\label{eq:dFdt}
    -\dot{F}_\psi (0)= - \la \psi | \cL (|\psi\ra\la\psi|) |\psi\ra=\gamma \frac{\Delta q^2_\psi+\Delta p^2_\psi - 1}{2}.
\end{equation}
The quadrature variances for the basis states are
\begin{eqnarray}
    \Delta q^2_{\bar0,\bar1}&=&\frac{\text{e}^{-2r}}{2\cN} \sum_{n,m=1}^N\left[ 1+\frac{\text{e}^{2r}}{2}(\bar{q}_n+\bar{q}_m)^2 \right] \\ 
    &&\times \exp \left[-\text{e}^{2r} \left( \frac{\bar{q}_n - \bar{q}_m}{2} \right)^2 \right], \nonumber \\
    \Delta p^2_{\bar0,\bar1}&=&\frac{\text{e}^{2r}}{2\cN} \sum_{n,m=1}^N\left[ 1-\frac{\text{e}^{2r}}{2}(\bar{q}_n - \bar{q}_m)^2  \right] \\
    &&\times \exp \left[-\text{e}^{2r} \left( \frac{\bar{q}_n - \bar{q}_m}{2} \right)^2 \right]. \nonumber
\end{eqnarray}

The growth of the initial fidelity decay rate~\eqref{eq:dFdt} with the position variance of the code states suggests that it becomes increasingly taxing to stabilize comb states with many teeth, as they exhibit a greater spread in phase space. In other words, the greater the average energy of a state $|\psi\ra$ compared to a vacuum state with the same average displacement, the faster fidelity decays. Indeed, we observe a quadratic growth of fidelity decay with the number of teeth.

Explicitly, the leading-order contribution to the above variances comes from the diagonal summands ($n=m$),
\begin{eqnarray}
    \Delta q_{\bar0,\bar1}^2 &\approx&\frac{\text{e}^{-2r}}{2}+\frac{1}{N} \sum_{n=1}^N\bar{q}_n^2=\frac{\text{e}^{-2r}}{2}+\frac{N^2-1}{12} d^2 
\end{eqnarray}
whereas $\Delta p^2_{\bar0,\bar1} \approx \text{e}^{2r}/2$. 
Off-diagonal terms are exponentially suppressed by $\exp(-\text{e}^{2r}d^2/4)$,
which leaves us with
\begin{equation}\label{eq:dFdt_approx}
    -\dot{F}_{\bar0,\bar1}(0) \approx \frac{\gamma}{2} \left(\frac{N^2-1}{12} d^2+\cosh 2r - 1\right).
\end{equation}
We note that the growing decay rate is \emph{not} related to the channel's actual damping of the coherent amplitude to zero. If we instead work with diffusion noise described by the generator~\eqref{eq:gaussianNoise}, the corresponding rate of fidelity decay would exhibit the same growth with the spread of the initial state in phase space,
\begin{eqnarray}\label{eq:dFtildedt_approx}
    -\dot{\tilde{F}}_\psi (0) &=&\gamma \left( \Delta q^2_\psi+\Delta p^2_\psi \right),  \\ 
    -\dot{\tilde{F}}_{\bar0,\bar1}(0) &\approx&\gamma \left(  \frac{N^2-1}{12} d^2+\cosh 2r \right). \nonumber
\end{eqnarray}

\begin{figure}
\includegraphics[width=0.95\linewidth]{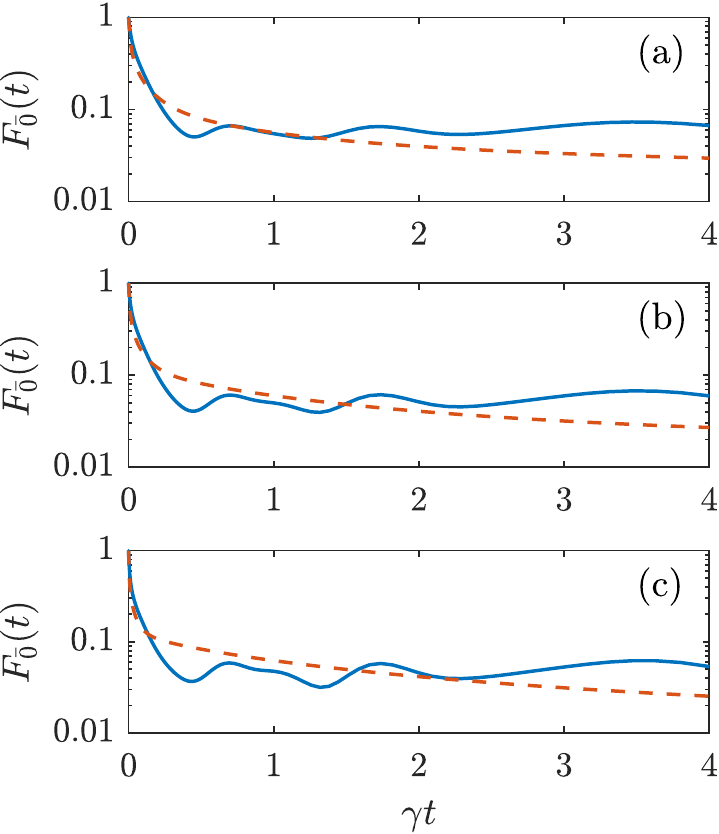} 
\caption{Variation of fidelity between the initial and final for the code states $\ket{\bar{0}}$ with time in the presence of a damping channel (blue) and a diffusion channel (red) for $N=8$ teeth. (a)-(c) Parameters $(d,r)=(4.0,0.5)$, $(5.0,0.3)$, and $(7.0,-0.1)$, respectively.}
\label{fig:F0plots}
\end{figure}

We plot the fidelity of a GKP basis state as a function of time in Fig.~\ref{fig:F0plots}, comparing the damping channel (solid lines) to the diffusion channel (dashed lines) for three exemplary parameter sets at $N=8$. They are chosen such that the error probability in distinguishing the basis states is less than $1\%$. After a fast initial decay, the damping channel leads to a fluctuating behavior as the $8$ displaced teeth slowly approach the vacuum state. The diffusion channel results in an initial decay at twice the rate, followed by a slow monotonous decrease as the comb diffuses. 

\begin{figure}
\includegraphics[width=0.95\linewidth]{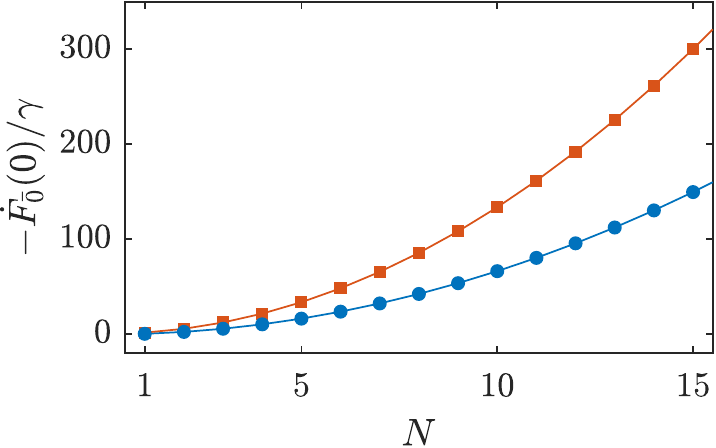} 
\caption{%
Derivative of fidelity between the initial and final code states $\ket{\bar0}$ at $t=0$ with $N$ in the presence of damping (blue) 
and diffusion (red) noise channels for $(d,r)=(4.0,0.5)$.
The corresponding solid lines indicate the scaling formul\ae~\eqref{eq:dFdt_approx} 
and~\eqref{eq:dFtildedt_approx}.%
}
\label{fig:dF0dt}
\end{figure}

The initial decay of fidelity is also shown in Fig.~\ref{fig:dF0dt} for the parameter set (a) at varying tooth number~$N$, comparing once again the damping channel (circles) to the diffusion channel (squares).
The latter case leads to an approximately doubled decay rate, which is also well described by the scaling formul\ae~\eqref{eq:dFdt_approx} and~\eqref{eq:dFtildedt_approx}. 
The observed growth with $N$ indicates the short-lived \emph{initial} code space spanned by large comb states. However, this does not equally determine the rate of noise-induced code errors, i.e.~the ability to decode the logical qubit from the noisy comb state.

\subsection{Orthogonality}
\label{subsec:orthogonality}

The growth of code errors over time in the presence of noise will manifest itself in a deteriorating orthogonality of the basis states. Consider the operator scalar product between the time-evolved mixed code states,
\begin{eqnarray}
    O(t) &=&\tr \left[ \rho_{\bar0}(t) \rho_{\bar1}(t) \right]=\tr \left[ \text{e}^{\cL t}(\ket{\bar0} \la \bar0|) \text{e}^{\cL t}(\ket{\bar1} \la \bar1|) \right] \nonumber \\
    &=& 2\pi \int \diff q \diff p \, w_{\bar0}(q,p;t) w_{\bar1}(q,p;t).
\end{eqnarray}
Its initial value $O(0)=\la \bar0\ket{\bar1}^2$ is determined by Eq.~\eqref{eq:orthog_pure}. Code errors are therefore already present in the absence of noise, but their probability grows with time. The damping channel, in particular, would drive the comb states towards $\vac$ and $O(\infty)=1$, i.e.~total loss of the logical qubit.

Again we focus on the rate of change relative to the initial value, which,
in the case of damping,
is
\begin{eqnarray}
    &&\dot{O}(0)=\gamma \left( \la \bar1 | \oa \ket{\bar0}^2+\la \bar0 | \oa \ket{\bar1}^2-2 \la \bar0\ket{\bar1} \la \bar0|\oa^\da \oa\ket{\bar1} \right) \\
    &&=\gamma \left( \la \bar0|\oq\ket{\bar1}^2 - \la \bar0|\op \ket{\bar1}^2+\la \bar0\ket{\bar1}^2 - \la \bar0\ket{\bar1} \la \bar0|\oq^2+\op^2 \ket{\bar1} \right). \nonumber
\end{eqnarray}
Here we use the fact that the wavefunctions $\la q|\bar0\rangle,\la q|\bar1\rangle\in\mathbb{R}$. Approximating the lengthy exact expression in the limit of non-overlapping teeth, we find to leading order,
\begin{eqnarray}
    -\frac{\dot{O}(0)}{O(0)}&\approx&\gamma \bigg\{ \cosh 2r - 1 + \frac{d^2}{4} \left[ \frac{N(N-1)}{12} \right. \nonumber \\ 
    & &\left.- e^{4r} \frac{2N^2-2N+1}{2(2N-1)^2} \right] \bigg\}. \label{eq:dO}
\end{eqnarray}
Notice that this rate of change is \emph{negative} for large combs ($N\gg1$), which implies that theoperator scalar product decreases and the basis states initially become more orthogonal. This is mainly due to the loss of purity: in terms of the operator scalar product, 
the two pure comb states $\ket{\bar0}$ and $\ket{\bar1}$ are less orthogonal  than the corresponding incoherent
mixtures of teeth. Specifically, we find that $O(0) = \la \bar 0 | \bar 1\ra^2$ exceeds the operator scalar product of the respective mixtures by the factor $(2N-1)$ in the limit of almost non-overlapping teeth. 

\begin{figure}
\includegraphics[width=0.95\linewidth]{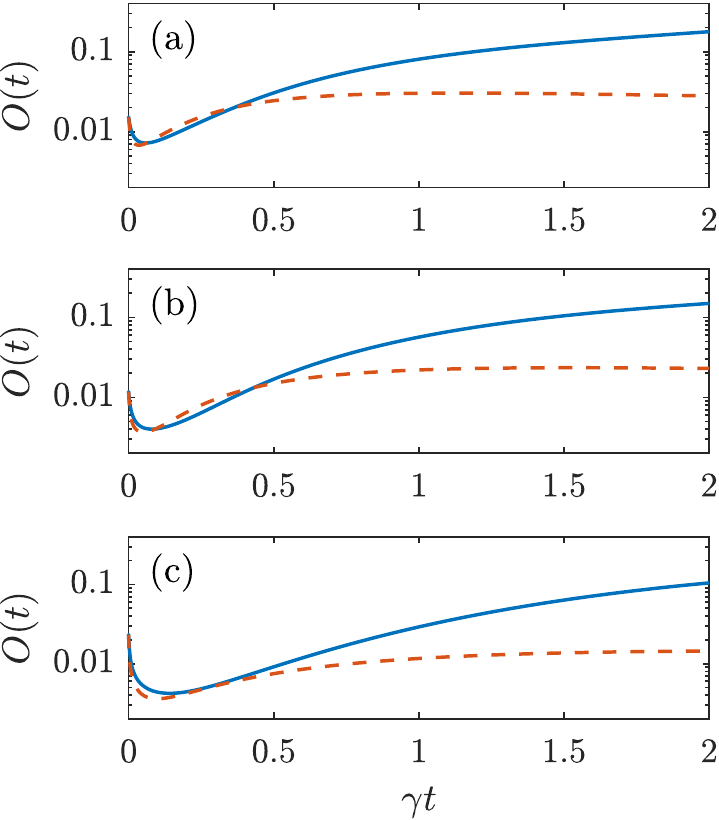} 
\caption{Variation of orthogonality between the code states $\ket{\bar{0}}$ and $\ket{\bar{1}}$ with time in the presence of the damping channel (blue) and diffusion channel (red) for the same parameter sets as in Fig.~\ref{fig:F0plots}.}
\label{fig:Oplots}
\end{figure}

For diffusion, we obtain
\begin{equation}
 \dot{\tilde{O}}(0)=2\gamma \left( \la \bar0|\oq\ket{\bar1}^2 - \la \bar0|\op \ket{\bar1}^2 - \la \bar0\ket{\bar1} \la \bar0|\oq^2+\op^2 \ket{\bar1} \right)
\label{eq:dOtilde} \end{equation}
and a corresponding leading-order approximation similar to Eq.~\eqref{eq:dO}. We compare the time dependence of the scalar product for damping and diffusion at the same three parameter settings as before in Fig.~\ref{fig:Oplots}.  The damping curves would eventually converge to unity, but only at much longer times. In all cases, the scalar product initially decreases and then increases again. 

\begin{figure}
\includegraphics[width=0.95\linewidth]{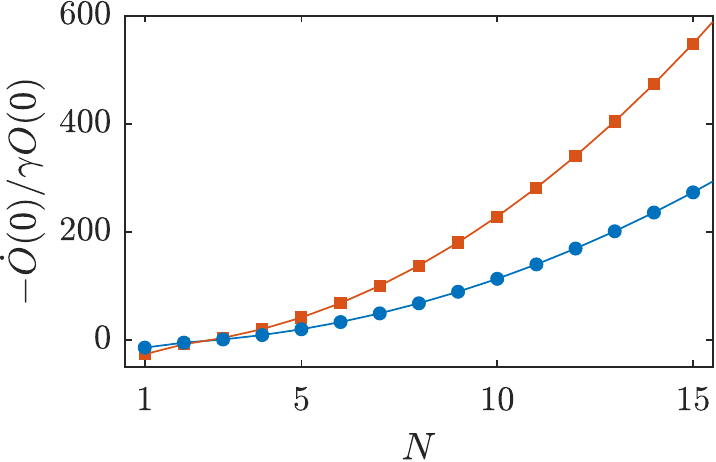} 
\caption{Derivative of orthogonality between the code states $\ket{\bar{0}}$ and $\ket{\bar{0}}$ at $t=0$ with~$N$ in the presence of damping (blue) and diffusion (red) noise channels for the same parameter values as in Fig.~\ref{fig:dF0dt}. The corresponding solid lines indicate the scaling formul\ae~\eqref{eq:dO} and~\eqref{eq:dOtilde}.}
\label{fig:dOdt}
\end{figure}

Figure \ref{fig:dOdt} shows the negative initial derivative as a function of~$N$, once again well approximated by the respective scaling formulas (solid lines). Any comb with more than two teeth results in an initially decreasing scalar product between the basis states. However, this must not be interpreted as a transient improvement of the encoding. As we will see in the following, code errors are related to the scalar product only in the case of pure code states.

\subsection{State distinguishability}
\label{subsec:distinguishability}

\begin{figure}
\includegraphics[width=0.95\linewidth]{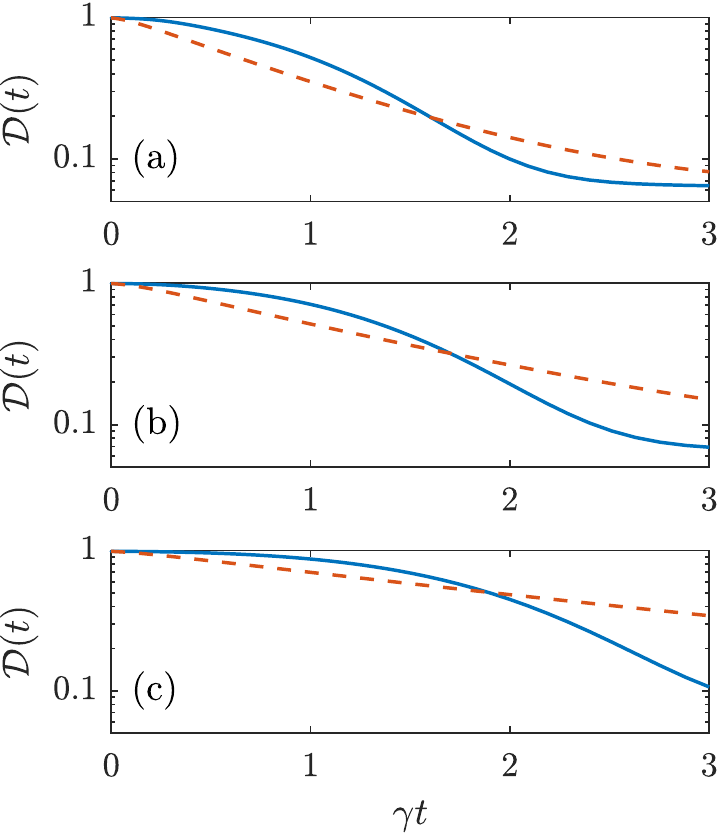}
\caption{Variation of state distinguishability with time in the presence of the damping channel (blue) and diffusion channel (red), using the same parameters as in Fig.~\ref{fig:F0plots}.}
\label{fig:Dplots}
\end{figure}

A reliable measure for code errors with a clear operational meaning is the state distinguishability.
Given two quantum states $\rho,\sigma$, 
state distinguishability is defined in terms of the trace distance \cite{Nelchu01},
\begin{equation}
     \cD (\rho,\sigma)  := \frac12 \tr |\rho-\sigma|=\frac12 \tr \sqrt{ (\rho-\sigma)^2 }.
\end{equation}
Quantum channels that describe continuous noise processes are contractive and thus imply a monotonous decay of $\cD$ for any pair of states.
The Holevo-Helstrom theorem~\cite{Helstrom76, Holevo73} states that the error probability for two-state discrimination by measurement is at least $\eps=(1-\cD)/2$. For the two initially pure basis states here, or for any two pure non-orthogonal states in fact, we can express the state distinguishability in terms of the scalar product as 
\begin{equation}
    \cD(\ket{\bar0}\la \bar0|,\ket{\bar1}\la \bar1|)=\sqrt{1 - |\la \bar0\ket{\bar1} |^2} \equiv \cD(0).
\end{equation}
Demanding a faithful initial encoding with a \emph{given} error bound
\begin{equation}
\frac{1-\cD(0)}{2}\leq\eps_{\max} \ll 1
\end{equation}
constrains the choice of comb parameters $(d,r,N)$ to the regime of almost non-overlapping teeth.
By virtue of Eq.~\eqref{eq:orthog_pure}, we then obtain the approximate constraint
\begin{equation}
    \eps\approx \frac{\la \bar{0}|\bar{1}\ra^2}{4} \approx \left( \frac{2N-1}{2N}\right)^2 \exp \left(-\frac{\text{e}^{2r}d^2}{8} \right) \leq \eps_{\max},
\end{equation}
which requires, in particular, a wide tooth spacing and/or strong squeezing with$e^r d\gg1$. The three exemplary configurations plotted in the three panels of Figs.~\ref{fig:F0plots}, \ref{fig:Oplots}, and~\ref{fig:Dplots} correspond to (a) $\eps \approx 0.4\%$, (b) $0.3\%$, and (c)~$0.6\%$. 

Under the influence of noise, we will have a lower distinguishability $\cD(t)$ of the two mixed code states $\rho_{\bar{0}} (t)$ and $\rho_{\bar{1}} (t)$ at a later point in time, 
\begin{equation}
    \cD(0) \geq \cD(t)=\frac12 \tr \sqrt{\left[ \rho_{\bar{0}}(t) - \rho_{\bar{1}}(t)  \right]^2} \xrightarrow{t\to \infty} 0.
\end{equation}
The values $\cD(t)$ for $t>0$ are no longer a simple expression of the operator scalar product; they must be evaluated by numerical diagonalization of the difference between the time-evolved code states in the square root. We can retrieve the corresponding density matrices in position representation from the time-evolved Wigner functions.
Given the damping channel~\eqref{eq:dissipation}, we arrive at
\begin{eqnarray}
    &&\left\la Q+\frac{q}{2} \right|\rho_{\bar{0}}(t)\left| Q - \frac{q}{2} \right\ra=\int \diff p\, w_{\bar{0}} \left( Q,p;t \right) \text{e}^{ipq} \\
    &=&\frac{1}{\sqrt{\pi} \cN\sigma_q(t)} \sum_{n,m=1}^N\exp \left\{ - \frac{\left[ 2Q-\text{e}^{-\gamma t/2}(\bar{q}_n+\bar{q}_m) \right]^2}{4\sigma_q^2 (t)} \right. \nonumber \\
    &&\left. - \frac{\sigma_p^2(t) q^2 + 2q(\bar{q}_n-\bar{q}_m)\text{e}^{2r-\gamma t/2}+ (\bar{q}_n-\bar{q}_m)^2 \text{e}^{2r} }{4} \right\}, \nonumber
\end{eqnarray}
and replacing $Q$ by $Q-\text{e}^{-\gamma t/2}d/2$ yields the density matrix for $\rho_{\bar{1}}(t)$. Alternatively, for the Gaussian diffusion channel~\eqref{eq:gaussianNoise}, we obtain
\begin{eqnarray}
    &&\left\la Q+\frac{q}{2} \right|\tilde{\rho}_{\bar{0}}(t)\left| Q - \frac{q}{2} \right\ra \\
    &=&\frac{1}{\sqrt{\pi} \cN\sqrt{\text{e}^{-2r} + 2\gamma t}} \sum_{n,m=1}^N\exp \left\{ - \frac{\left( \bar{q}_n+\bar{q}_m -2Q \right)^2}{4(\text{e}^{-2r} + 2\gamma t)} \right. \nonumber \\
    &&\left. - \frac{2\gamma t q^2+\text{e}^{2r}(\bar{q}_n - \bar{q}_m + q)^2 }{4} \right\}. \nonumber
\end{eqnarray}
Taking only the diagonal terms in the double sum, we can evaluate the distinguishability of the incoherent counterparts to the comb states $|\bar 0\ra$ and $|\bar 1\ra$. Remarkably, and in contrast to the discrepancy of the scalar products in \S\ref{subsec:orthogonality}, we find that incoherent combs are as distinguishable as coherent ones at initial time, within numerical accuracy in the three parameter cases considered.

Variation of the distinguishability with time is plotted in Fig.~\ref{fig:Dplots} for the damping (solid line) and the diffusion (dashed line) channels. We observe a monotonous behaviour in both cases, but surprisingly, the initial slope in the diffusion case is significantly (and not just by the factor of two) steeper than in the damping case. The detrimental influence of damping exceeds that of the diffusion channel only at a later point in time, when the accumulated code error is no longer tenable. 

\begin{figure}
\includegraphics[width=0.95\linewidth]{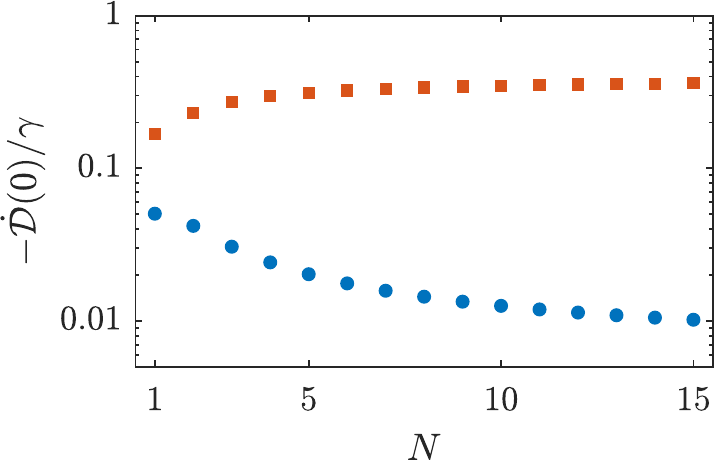}
\caption{Derivative of state distinguishability at $t=0$ with~$N$ in the presence of damping (blue) and diffusion (red) noise channels for the same parameters as in Fig.~\ref{fig:dF0dt}}
\label{fig:dDdt}
\end{figure}

To clarify this further, we also plot the negative derivative of the distinguishability at time $t=0$ as a function of~$N$ in Fig.~\ref{fig:dDdt}. In the limiting case $N=1$ of a coherent-state encoding, the difference between damping (circles) and diffusion (squares) is indeed roughly a factor two. With growing~$N$ however, the disparity quickly rises to more than an order of magnitude, highlighting a strikingly different sensitivity of GKP codes to damping and diffusion---a property that does not reflect in the analytically tractable scaling behaviour of initial-state fidelity and orthogonality. In fact, the opposite scaling of the distinguishability with~$N$ suggests that it is detrimental to compensate the systematic effect of damping by linear amplification when implementing error correction on GKP codes. Apparently, it is easier to stabilize comb states under the damping channel than under the diffusion channel.

\section{Conclusion}\label{sec:conclusions}

We have studied properties of squeezed comb states, a realistic implementation of GKP encoding using superpositions of a finite number of squeezed coherent states arranged equidistantly along the position quadrature axis. With the help of the Wigner function representation, we have characterized the peculiar interference features of these states, as well as their time evolution under the influence of two important noise processes: amplitude damping and Gaussian diffusion. This dynamical phase-space framework can help to clarify the noise sensitivity of GKP encoding as a function of its constituting parameters: the number of comb teeth, the spacing, and the squeezing. To this end, we have assessed the behaviour of several figures of merit for the stability of GKP encoding in the presence of noise.

Specifically, we have evaluated distinguishability between encoded computational basis states subjected to damping and diffusion noise, which directly measures the susceptibility to code errors. We have found that GKP states are substantially more robust against the
initial buildup of errors due to amplitude damping (pure loss) than due to diffusion. The discrepancy grows with the comb size, and it suggests that one should avoid error correction strategies based on the conversion of damping to diffusion noise by means of linear amplification~\cite{Albert18}. 
An alternative scheme to stabilize GKP states against amplitude damping is proposed in Ref.~\cite{Royer2020}.
Our phase-space approach based on the explicit time evolution of GKP states subject to noise puts a spotlight on the dynamical description of code errors and, potentially, continuous online protocols for code stabilization.

\acknowledgments
N.S.\ acknowledges the Max Planck Institute for the Science of Light for the financial support of the postdoctoral fellowship.
B.C.S.\ acknowledges NSERC for financial support.
B.C.S.\ and N.S.\ acknowledge the traditional owners of the land on which some of this work was undertaken at the University of Calgary: the Treaty 7 First Nations.


%

\end{document}